\documentclass[sigconf,nonacm,natbib=false]{acmart}
\makeatletter
\global\@ACM@balancefalse
\makeatother

\usepackage{amsmath}
\usepackage{algorithm}
\usepackage{algpseudocode}
\usepackage{array}
\usepackage{booktabs}
\usepackage{multirow}
\usepackage{placeins}
\usepackage{cuted}

\setcopyright{none}
\renewcommand\footnotetextcopyrightpermission[1]{}
\usepackage[
  backend=biber,
  style=numeric,
  sorting=none,
  giveninits=true,
  maxbibnames=3,
  minbibnames=3,
  doi=false,
  isbn=false,
  url=false,
  eprint=false
]{biblatex}
\DeclareNameAlias{author}{family-given}
\DeclareNameAlias{editor}{family-given}

\renewrobustcmd*{\bibinitperiod}{}
\renewrobustcmd*{\bibinitdelim}{}
\DeclareDelimFormat{multinamedelim}{\addcomma\space}
\DeclareDelimFormat{finalnamedelim}{\addcomma\space}
\DefineBibliographyStrings{english}{andothers={et\addabbrvspace al\adddot}}
\DeclareFieldFormat[article,inproceedings,incollection,misc]{title}{#1}
\DeclareFieldFormat{pages}{#1}

\AtBeginBibliography{\sloppy}

\newbibmacro*{compact:author-title}{%
  \printnames{author}%
  \setunit{\addperiod\space}%
  \printfield{title}}

\DeclareBibliographyDriver{article}{%
  \usebibmacro{bibindex}\usebibmacro{begentry}%
  \usebibmacro{compact:author-title}\newunit\newblock
  \printfield{journaltitle}\setunit{\addcomma\space}\printfield{year}%
  \iffieldundef{volume}{}{\setunit{\addcomma\space}\printfield{volume}%
    \iffieldundef{number}{}{\printtext{\mkbibparens{\printfield{number}}}}}%
  \iffieldundef{pages}{}{\setunit{\addcolon\space}\printfield{pages}}%
  \usebibmacro{finentry}}

\DeclareBibliographyDriver{inproceedings}{%
  \usebibmacro{bibindex}\usebibmacro{begentry}%
  \usebibmacro{compact:author-title}\newunit\newblock
  \printfield{booktitle}\setunit{\addcomma\space}\printfield{year}%
  \iffieldundef{pages}{}{\setunit{\addcolon\space}\printfield{pages}}%
  \usebibmacro{finentry}}

\DeclareBibliographyDriver{misc}{%
  \usebibmacro{bibindex}\usebibmacro{begentry}%
  \usebibmacro{compact:author-title}%
  \iffieldundef{howpublished}{}{\setunit{\addperiod\space}\printfield{howpublished}}%
  \iffieldundef{year}{}{\setunit{\addcomma\space}\printfield{year}}%
  \usebibmacro{finentry}}

\title{Research Team Identification Based on Representation Learning of Academic Heterogeneous Information Network}

\author{Junfu Wang}
\affiliation{%
  \institution{Beijing Key Laboratory of Intelligent Communication Software and Multimedia, School of Computer Science (National Pilot Software Engineering School), Beijing University of Posts and Telecommunications}
  \city{Beijing}
  \country{China}}

\author{Yawen Li}
\authornote{Corresponding author: warmly0716@126.com.}
\affiliation{%
  \institution{School of Economics and Management, Beijing University of Posts and Telecommunications}
  \city{Beijing}
  \country{China}}

\author{Zhe Xue}
\affiliation{%
  \institution{Beijing Key Laboratory of Intelligent Communication Software and Multimedia, School of Computer Science (National Pilot Software Engineering School), Beijing University of Posts and Telecommunications}
  \city{Beijing}
  \country{China}}

\author{Ang Li}
\affiliation{%
  \institution{Beijing Key Laboratory of Intelligent Communication Software and Multimedia, School of Computer Science (National Pilot Software Engineering School), Beijing University of Posts and Telecommunications}
  \city{Beijing}
  \country{China}}

\begin{abstract}
Academic networks in the real world can usually be described by heterogeneous information networks composed of multiple types of nodes and relationships. Existing representation-learning research for homogeneous information networks lacks the ability to explore the heterogeneity of such networks and therefore cannot be directly applied to heterogeneous information networks. To meet the practical need to identify and discover scientific research teams from academic heterogeneous information networks composed of massive and complex scientific and technological data, this paper proposes a research-team identification method based on representation learning. Node-level and meta-path-level attention mechanisms learn low-dimensional, dense, real-valued vector representations while retaining rich topological information and meta-path semantics. Scientific research teams and important team members are then identified by maximizing node influence. Experimental results show that the proposed method outperforms the comparison methods.
\end{abstract}


\begin{document}
\maketitle

\section{Introduction}

Scientific and technological data, including journal papers, funded projects, and patents, are growing rapidly. Ontology-based retrieval has long supported the organization and access of such information~\cite{yang2015ontology}, while studies of scientific collaboration describe how research relationships evolve in the information age~\cite{yao2020globalVillage}. These multi-type, multi-form, and widely connected data constitute large academic heterogeneous information networks (HINs), including social platforms whose heterogeneous relations support personalized recommendation~\cite{elKishky2022twhin}. Dynamic-interest tracking can further reveal changing scholar groups from multiple views~\cite{li2023scholarClustering}, and sentiment-variation modeling shows how temporal changes in large public-event data can be explained~\cite{li2025sentimentSpike}. Because scientific work is increasingly specialized and complex, teamwork has become an important way to advance research. Effectively discovering research teams from large and complex scientific data is therefore an urgent practical need.

Network embedding preserves proximity and semantic information in large networks. Low-cost incremental learning has been studied for dynamic HINs~\cite{peng2021lime}, and semantic-similarity attention combined with hypergraph convolution strengthens scientific-publication representations~\cite{li2026semanticHypergraph}. HIN embedding has been applied to clustering, classification, link prediction, and recommendation. Relation-structure-aware embedding captures heterogeneous relations~\cite{lu2019relationStructure}; interpretable machine learning can expose the evidence behind intelligent decisions~\cite{li2019interpretableDecision}; and graph neural network surveys summarize the broader development of representation learning on graphs~\cite{wu2020gnnSurvey}. However, many meta-path-based approaches assume that all nodes share the same meta-path weights and therefore cannot express personalized semantic preferences.

Scientific and technological information is often multimodal. Semantics-adversarial and media-adversarial learning improves cross-media retrieval under heterogeneous information sources~\cite{li2022crossMediaAdversarial}, while retrieval-oriented masked autoencoding learns stronger text representations~\cite{xiao2022retroMae}. Heterogeneous graph attention networks support semi-supervised short-text classification~\cite{hu2019heterogeneousGraphAttention}, and deep modularity learning provides a complementary view of community discovery~\cite{yang2016deepModularity}. These developments motivate a representation model that jointly preserves network topology and meta-path semantics.

To address this challenge, we propose an attention-based representation-learning method for academic HINs and a node-influence-maximization method for research-team discovery. The main contributions are as follows:

\begin{itemize}
  \item We construct an academic HIN that represents multiple types of scientific and technological data with complex structures and implicit semantics.
  \item We propose a node-level and meta-path-level attention model that learns low-dimensional, dense, real-valued representations of academic HIN nodes.
  \item We propose a research-team identification method that discovers teams, leaders, and important members using learned representations and node influence.
\end{itemize}

\section{Related Work}

HINs are an effective paradigm for modeling complex relations and objects. Relation-aware modeling remains important when node and edge types have distinct meanings~\cite{lu2019relationStructure}. GraphGAN uses adversarial learning to fit the distribution of links between nodes~\cite{wang2018graphgan}, whereas HIN-DRL embeds different node types through dynamic random walks~\cite{lu2020hindrl}. Dynamic meta-path proximity further supports evolving HIN embeddings~\cite{wang2020dynamicHin}. Methods for incomplete multi-view semi-supervised classification address missing views~\cite{xue2019incomplete}, and T2-GNN uses teacher--student distillation when graph features and structure are incomplete~\cite{huo2023t2gnn}. Self-supervised graph co-training can also exploit complementary relational signals~\cite{xia2021graphCoTraining}.

Existing research-team identification methods can be divided into traditional approaches, association-rule approaches, hierarchical clustering, and social-network analysis. Topic-trend prediction combines recurrent and graph models to analyze scientific research evolution~\cite{xu2022trend}. Association-rule mining can identify research and development teams from patent data~\cite{lv2016teamIdentification}; hierarchical clustering can detect edge-weighted communities~\cite{li2019community}; and iterative betweenness-centrality ranking can identify team leaders~\cite{yu2018teamSocial}. Research on technology-cooperation networks further explains team formation mechanisms~\cite{ma2022cooperation}. These approaches either depend strongly on institutional or project information, do not identify non-core members, or cannot adequately express node heterogeneity.

Federated representation learning offers another way to coordinate heterogeneous data owners. Federated graph neural networks support cross-graph node classification~\cite{guan2021federatedGnn}, and FedSIN learns information-network representations through federated self-adaptation~\cite{li2026fedSin}. Communication-efficient reinforcement federated learning combines dynamic client selection with adaptive gradient compression~\cite{pan2025rfcsc}, while reinforcement-based active client selection targets heterogeneous graph learning~\cite{wang2025activeClientSelection}. Federated supervised cross-modal retrieval learns shared retrieval spaces without centralizing modality-specific data~\cite{li2024federatedCrossModal}. Although the present work focuses on a unified academic HIN, these studies illustrate the value of robust representation learning across heterogeneous sources.

Other related advances include InfoMax-enhanced few-shot node classification~\cite{xu2023infomax}, selective reinforced sequence-to-sequence attention for social-media summarization~\cite{liang2020summarization}, multi-view neural prediction~\cite{li2022fuelConsumption}, and attention-based stock-trend prediction~\cite{chen2020stock}. Learnable-edge collaborative filtering~\cite{xiao2022lecf}, multi-feature hashtag recommendation~\cite{kou2018hashtag}, and deep collaborative filtering with multi-aspect information~\cite{shi2019deepCollaborative} demonstrate the breadth of heterogeneous recommendation tasks. Filter-enhanced multilayer perceptrons provide an efficient alternative for sequential recommendation~\cite{zhou2022filterMlp}, and bi-projection fusion integrates complementary views for omnidirectional image super-resolution~\cite{wang2024omnidirectionalSr}. Even developments outside network analysis, such as thermochromic cluster materials~\cite{xue2021luminescent} and multiagent tracking~\cite{meng2013tracking}, illustrate the diverse scientific records that an academic information network may need to organize.

\section{Proposed Method}

\begin{figure*}[t]
  \centering
  \includegraphics[width=\textwidth]{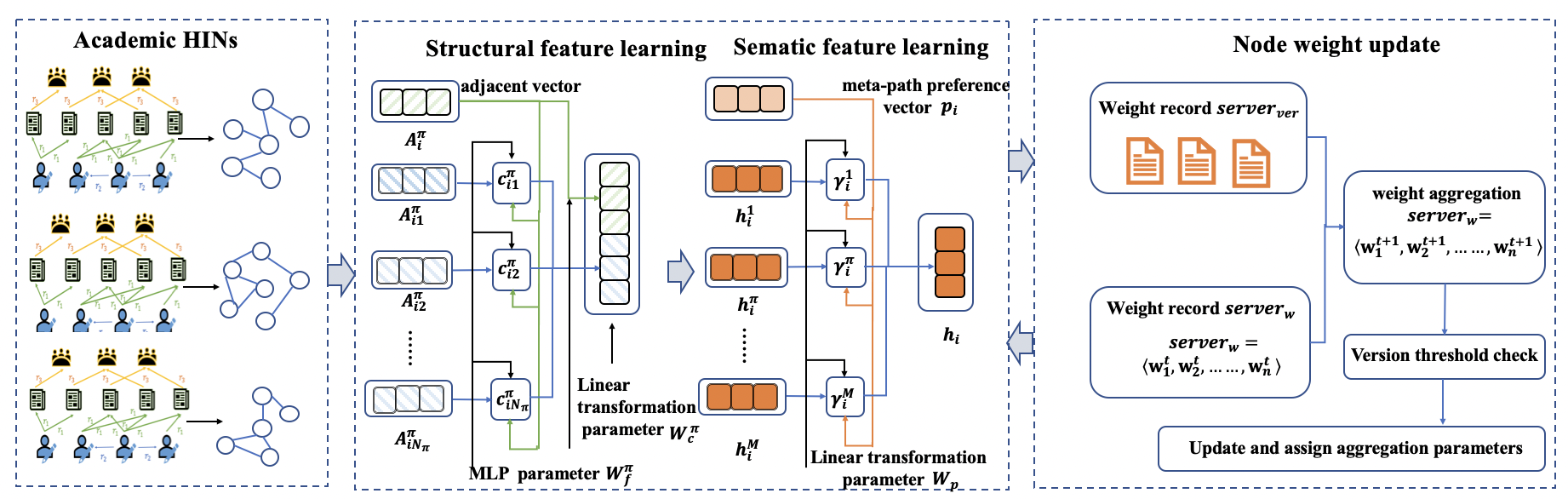}
  \caption{Architecture of the Academic Heterogeneous Information Network Representation Learning Model (AHinE).}
  \label{fig:architecture}
\end{figure*}

The architecture of the Academic Heterogeneous Information Network Representation Learning Model (AHinE) is shown in Figure~\ref{fig:architecture}. Node-level attention learns structural features by mining topology, and meta-path-level attention learns semantic features from heterogeneous relations. During learning, node weights are dynamically aggregated and updated using version and weight records.

\subsection{Structural Feature Representation Learning for Academic HINs}

The node-level attention mechanism uses normalized meta-path adjacency vectors to learn structural representations by aggregating a node and its neighbors connected through path instances. Because a meta-path adjacency vector is high-dimensional and sparse, a multilayer perceptron maps it into a $d$-dimensional space. For node $v_i$ and a neighbor $v_j$ connected by meta-path $\pi$, their structural similarity is

\begin{equation}
c_{ij}^{\pi}=\frac{(W_f^{\pi}A_i^{\pi})^{\mathsf T}W_f^{\pi}A_j^{\pi}}
{\lVert W_f^{\pi}A_i^{\pi}\rVert\,\lVert W_f^{\pi}A_j^{\pi}\rVert}.
\end{equation}

The normalized node-level attention coefficient is

\begin{equation}
s_{ij}^{\pi}=\frac{\exp(c_{ij}^{\pi})}
{\sum_{v_k\in\mathcal N_i^{\pi}}\exp(c_{ik}^{\pi})}.
\end{equation}

The aggregated neighbor representation and the resulting meta-path representation are

\begin{align}
h_{\mathcal N_i}^{\pi}&=\sigma\!\left(\sum_{v_j\in\mathcal N_i^{\pi}}s_{ij}^{\pi}W_f^{\pi}A_j^{\pi}\right),\\
h_i^{\pi}&=W_c^{\pi}\big[h_{\mathcal N_i}^{\pi};W_f^{\pi}A_i^{\pi}\big],
\end{align}

where $W_f^{\pi}$ and $W_c^{\pi}$ are trainable transformations, $A_i^{\pi}$ is the normalized meta-path adjacency vector, $\mathcal N_i^{\pi}$ is the meta-path neighborhood, and $[\,;\,]$ denotes concatenation.

\subsection{Semantic Feature Representation Learning for Academic HINs}

After learning structural representations, meta-path-level attention captures the different semantic contributions of meta-paths. A nonlinear transformation first maps the meta-path representation into a $k$-dimensional space:

\begin{equation}
h_i^{\prime\pi}=\sigma(W_p^{\pi}h_i^{\pi}+b_p^{\pi}).
\end{equation}

With a node-specific meta-path preference vector $p_i^{\pi}$, the compatibility score and its normalized attention weight are

\begin{align}
g_i^{\pi}&=(p_i^{\pi})^{\mathsf T}h_i^{\prime\pi},\\
\gamma_i^{\pi}&=\frac{\exp(g_i^{\pi})}{\sum_{m=1}^{M}\exp(g_i^{m})}.
\end{align}

The final representation of node $v_i$ is

\begin{equation}
h_i=\sum_{\pi=1}^{M}\gamma_i^{\pi}h_i^{\pi}.
\end{equation}

When some nodes are labeled, the model is trained semi-supervisedly with mini-batch stochastic gradient descent and backpropagation. The objective minimizes cross-entropy:

\begin{equation}
\mathcal L=-\sum_i\sum_l Y_{il}\log \widehat{Y}_{il},
\end{equation}

where $Y_{il}$ and $\widehat{Y}_{il}$ denote the true and predicted values for label $l$.

\subsection{Node Weight Update for Academic HINs}

The parameter server maintains the latest weight and version of every working node. Weighted aggregation removes obsolete updates from slow nodes while allowing fast nodes to update promptly. Let $w_i$ be the latest model from worker $i$, $v_i$ its version, $v_{\mathrm{latest}}$ the latest server version, and $\alpha$ the version-decay factor. The aggregation is

\begin{equation}
w_{\mathrm{server}}\leftarrow
\frac{\sum_{i=1}^{n}\alpha^{v_{\mathrm{latest}}-v_i}w_i}
{\sum_{i=1}^{n}\alpha^{v_{\mathrm{latest}}-v_i}}.
\end{equation}

Thus, the normalized worker weight is

\begin{equation}
\bar\alpha_i=\frac{\alpha^{v_{\mathrm{latest}}-v_i}}
{\sum_{j=1}^{n}\alpha^{v_{\mathrm{latest}}-v_j}}.
\end{equation}

If a worker exceeds a version-gap threshold, the server sends it the latest global weights; otherwise only the updated aggregation parameters are returned. This reduces the negative effect of different local training speeds.

\subsection{Research Team Identification}

AHinE representations combine neighbor-based topology and meta-path semantics. For nodes connected through meta-path instances, the node-level coefficient $s_{ij}^{\pi}$ measures neighbor influence. The average attention received from neighbors represents node importance. The most influential author is selected as a team leader.

Authors in the same team usually have close co-authorship or affiliation relations and therefore similar learned representations. The top-$K$ authors most similar to the leader are taken as member candidates. Their intersection with the leader's top-$K'$ influential neighbors gives the core members; remaining direct author neighbors are treated as non-core members. The procedure repeats on unassigned authors, as shown in Algorithm~\ref{alg:ahinrti}.

\begin{algorithm}[t]
\caption{AHinRTI}
\label{alg:ahinrti}
\begin{algorithmic}[1]
\Require Academic heterogeneous information network
\Ensure Scientific research-team identification results
\State Obtain vector representations of network nodes using AHinE
\ForAll{unrecognized author nodes}
  \State Compute and rank the influence of each unmarked author
  \State Select the top-influence author as the current team leader and mark it recognized
  \State Select the $K$ authors most similar to the leader as core-member candidates
  \State Intersect the candidates with the leader's top-$K'$ influential neighbors
  \State Mark the intersection as core members
  \State Mark the remaining neighboring authors as non-core members
  \State Output the current team and its member roles
\EndFor
\State \Return all identified teams
\end{algorithmic}
\end{algorithm}

Normalized Mutual Information (NMI) evaluates agreement between identified and actual teams:

\begin{equation}
\mathrm{NMI}=\frac{-2\sum_{i=1}^{C_1}\sum_{j=1}^{C_2}M_{ij}
\log\!\left(\frac{M_{ij}N}{M_{i\cdot}M_{\cdot j}}\right)}
{\sum_{i=1}^{C_1}M_{i\cdot}\log(M_{i\cdot}/N)+
 \sum_{j=1}^{C_2}M_{\cdot j}\log(M_{\cdot j}/N)},
\end{equation}

where $C_1$ and $C_2$ are the actual and identified numbers of teams, $M$ is their confusion matrix, and $N$ is the number of evaluated nodes.

\begin{table*}[t]
\caption{Node classification on three academic HIN datasets.}
\label{tab:classification}
\centering
\begin{tabular*}{\textwidth}{@{\extracolsep{\fill}}l*{6}{c}}
\toprule
\multirow{2}{*}{Method} & \multicolumn{2}{c}{DBLP} & \multicolumn{2}{c}{AMiner} & \multicolumn{2}{c}{ACM} \\
\cmidrule(lr){2-3}\cmidrule(lr){4-5}\cmidrule(lr){6-7}
& Micro-F1 & Macro-F1 & Micro-F1 & Macro-F1 & Micro-F1 & Macro-F1 \\
\midrule
Node2Vec     & 0.836 & 0.837 & 0.841 & 0.826 & 0.874 & 0.865 \\
LINE         & 0.841 & 0.843 & 0.899 & 0.881 & 0.913 & 0.896 \\
MetaPath2Vec & 0.783 & 0.776 & 0.923 & 0.878 & 0.921 & 0.887 \\
HIN2Vec      & 0.850 & 0.852 & 0.927 & 0.902 & 0.927 & 0.904 \\
GAT          & 0.843 & 0.840 & 0.921 & 0.899 & 0.867 & 0.876 \\
AHinE        & \textbf{0.867} & \textbf{0.859} & \textbf{0.932} & 0.901 & \textbf{0.931} & \textbf{0.913} \\
\bottomrule
\end{tabular*}
\end{table*}

\section{Experiments}

\subsection{Node Classification}

Macro-F1 and Micro-F1 evaluate node-classification performance on DBLP, AMiner, and ACM. Table~\ref{tab:classification} reports the results. AHinE outperforms homogeneous network embeddings Node2Vec and LINE on all metrics, indicating that it captures heterogeneous information effectively. Compared with MetaPath2Vec and HIN2Vec, which do not learn personalized weights for different meta-paths, AHinE achieves the best or near-best results. Its advantage over GAT further supports the effectiveness of the proposed node-level and meta-path-level attention mechanisms.

\begin{figure*}[t]
  \centering
  \includegraphics[width=0.9\textwidth]{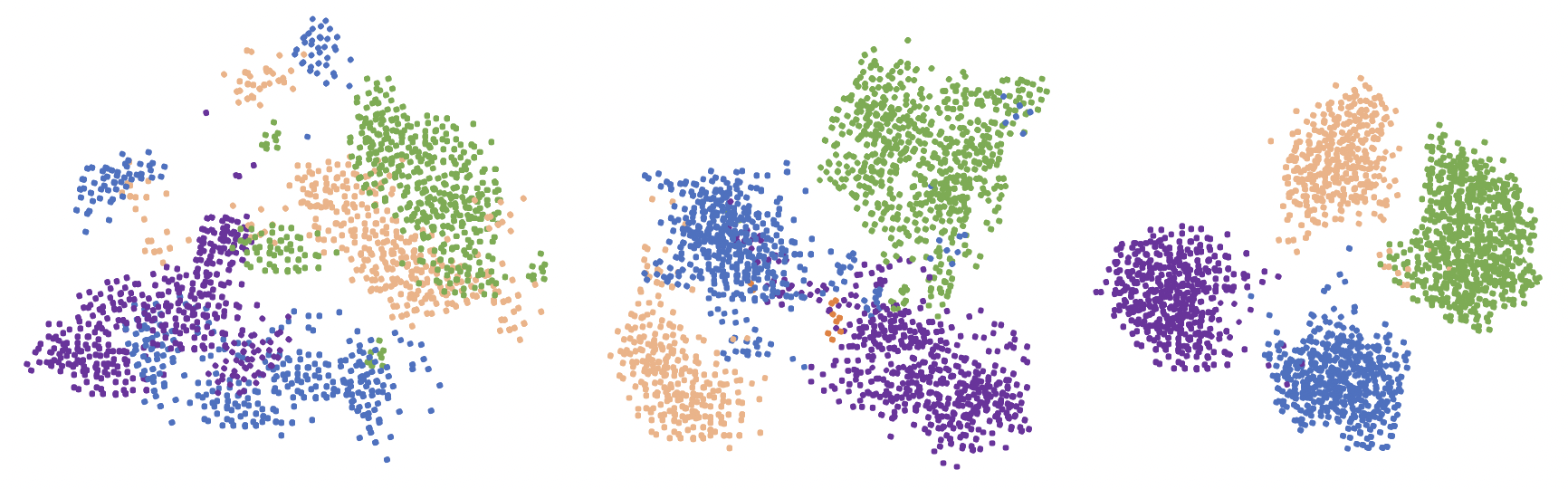}
  \caption{Network visualization results on the DBLP dataset with three embedding methods: (a) MetaPath2Vec, (b) HIN2Vec, and (c) AHinE.}
  \label{fig:visualization}
\end{figure*}

Figure~\ref{fig:visualization} visualizes DBLP embeddings after t-SNE dimensionality reduction. AHinE produces more compact clusters and clearer boundaries than MetaPath2Vec and HIN2Vec, showing that it learns topology and semantics more effectively.

\subsection{Parameter Sensitivity Analysis}

\begin{figure*}[t]
  \centering
  \begin{minipage}[t]{0.48\textwidth}
    \centering
    \includegraphics[width=\linewidth]{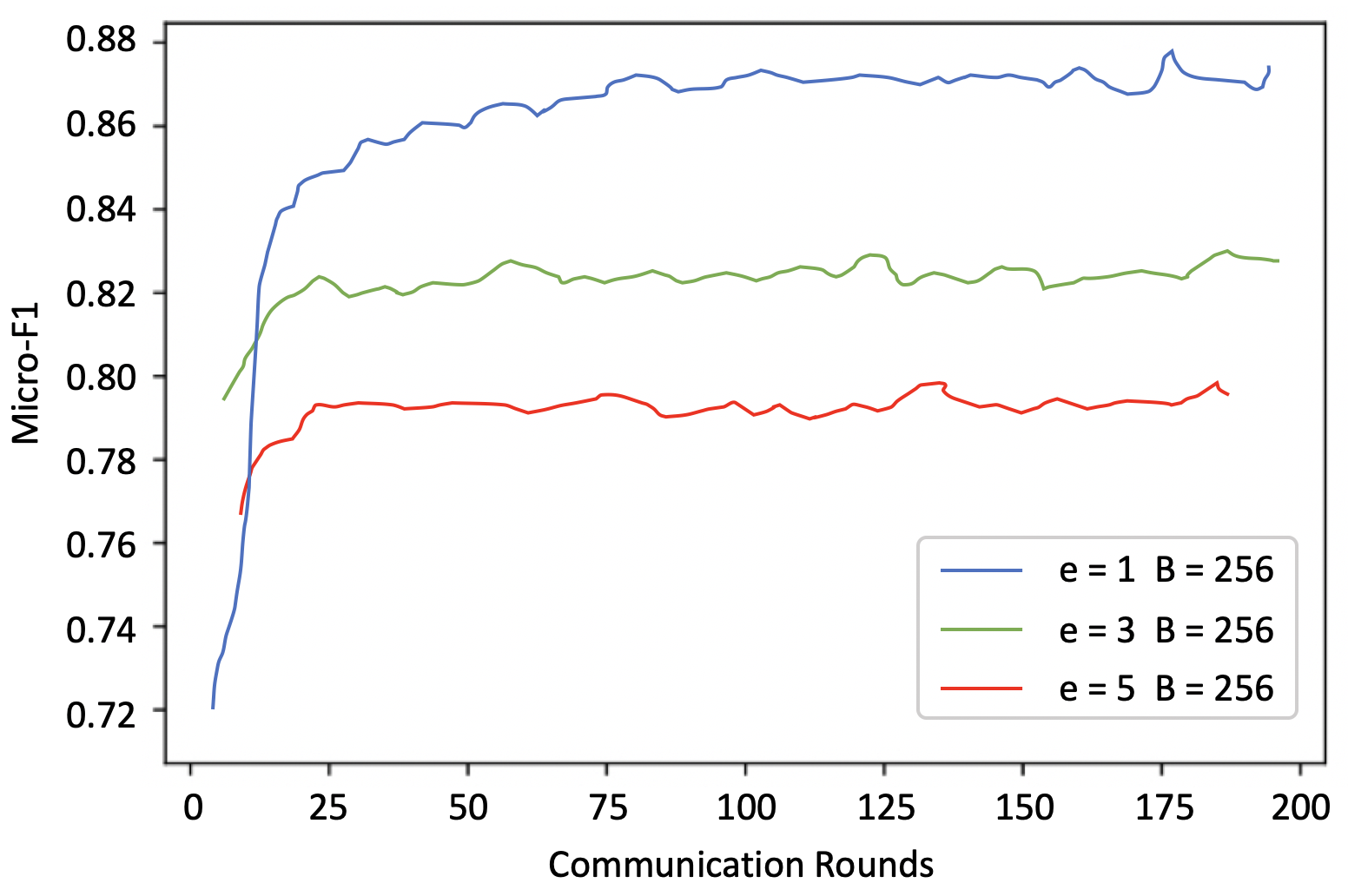}
    \small (a) Fixed batch size with varying local epochs.
  \end{minipage}\hfill
  \begin{minipage}[t]{0.48\textwidth}
    \centering
    \includegraphics[width=\linewidth]{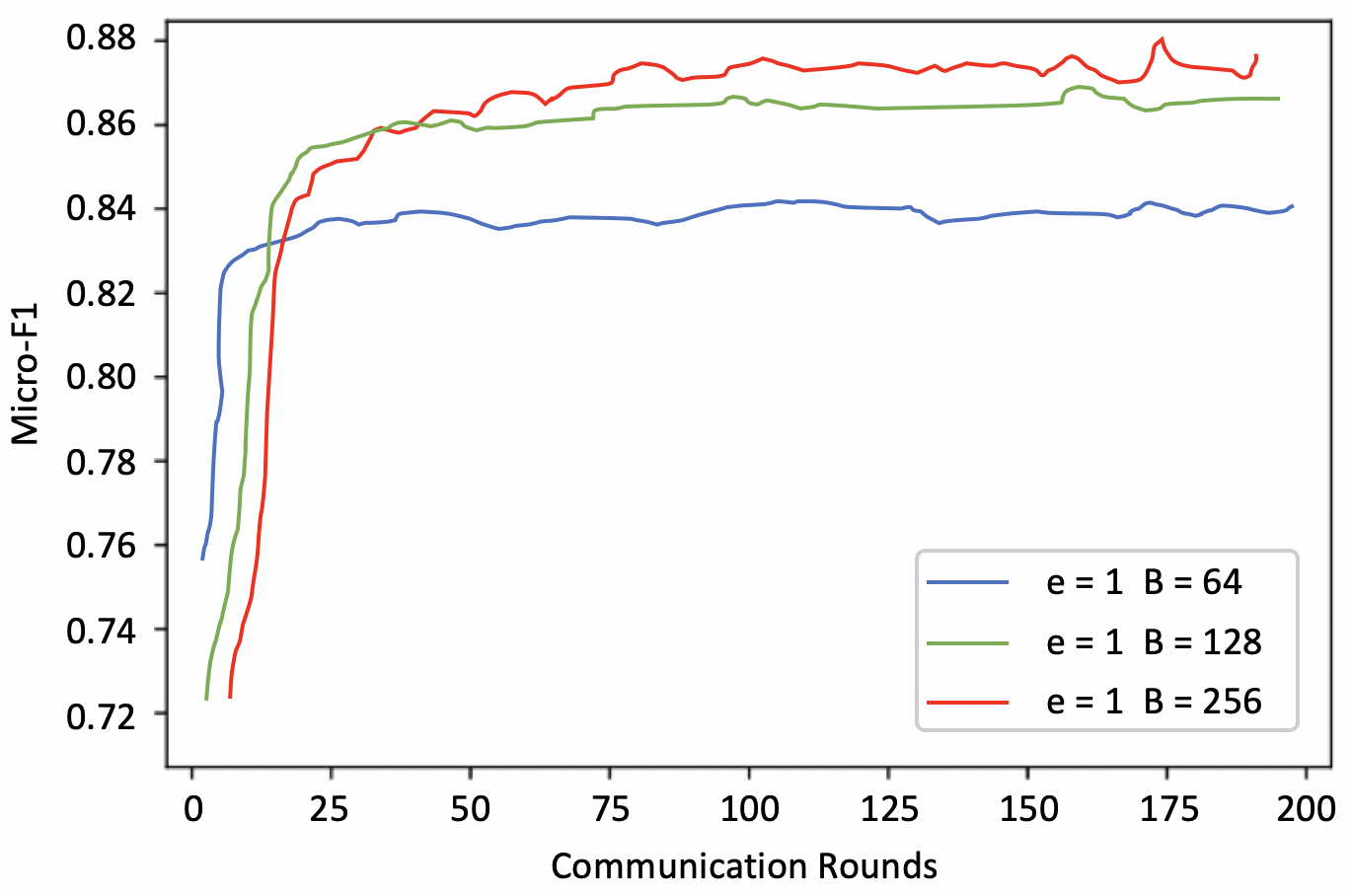}
    \small (b) Fixed local epochs with varying batch sizes.
  \end{minipage}
  \caption{Results of parameter sensitivity analysis on DBLP.}
  \label{fig:sensitivity}
\end{figure*}

We examine the effect of client computation on convergence. Client computation is represented by $e/B$, where $e$ is the local epoch number and $B$ is the batch size. On DBLP, the client number is fixed to three, local epochs are selected from $\{1,3,5\}$, and batch sizes from $\{64,128,256\}$. Figure~\ref{fig:sensitivity} shows that fewer local epochs can improve final performance but slow convergence, whereas larger batches improve performance while reducing convergence speed. AHinE performs best with $e=1$ and $B=256$.

\subsection{Node Influence Evaluation}

\begin{figure*}[t]
  \centering
  \includegraphics[width=0.58\textwidth]{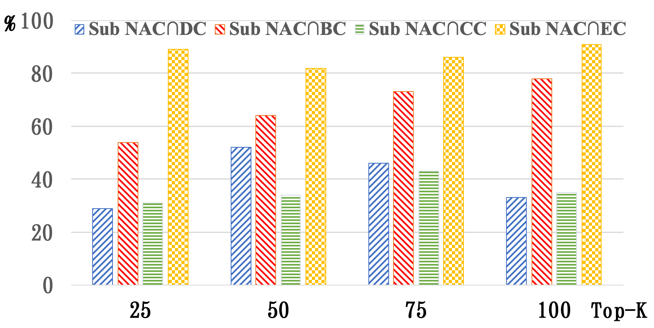}
  \caption{Node importance evaluation results on DBLP.}
  \label{fig:dblp-importance}
\end{figure*}

\begin{figure*}[t]
  \centering
  \includegraphics[width=0.58\textwidth]{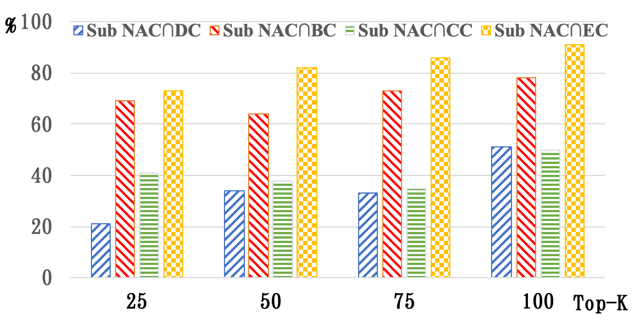}
  \caption{Node importance evaluation results on AMiner.}
  \label{fig:aminer-importance}
\end{figure*}

\begin{figure*}[t]
  \centering
  \includegraphics[width=0.58\textwidth]{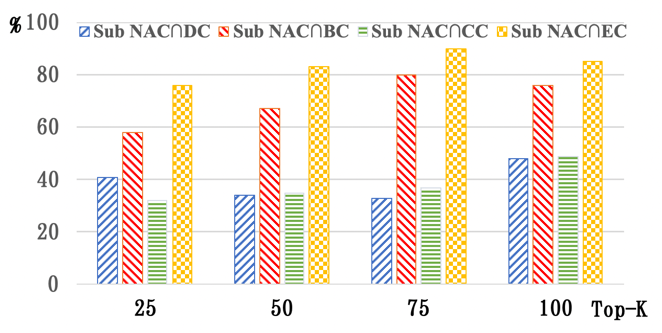}
  \caption{Node importance evaluation results on ACM.}
  \label{fig:acm-importance}
\end{figure*}

The node-level attention coefficient (NAC) evaluates author-node importance. We compare the top-$K$ NAC nodes with the top-$K$ nodes from degree centrality (DC), betweenness centrality (BC), closeness centrality (CC), and eigenvector centrality (EC). Figures~\ref{fig:dblp-importance}--\ref{fig:acm-importance} report the intersection percentages on DBLP, AMiner, and ACM. NAC agrees strongly and consistently with BC and EC. Its lower and less stable overlap with DC and CC reflects the fact that scientific-team importance depends on both a node and the importance of its neighbors, rather than only degree or geometric centrality.

\subsection{Research Team Identification}

We measure author similarity and node influence using the learned representations. To obtain reliable teams, the experiment retains authors with at least 10 publications and co-authorship frequency of at least five. Table~\ref{tab:teams} shows the top five teams identified on DBLP.

\begin{strip}
\centering
\captionof{table}{Identification results for the top five scientific research teams.}
\label{tab:teams}
\begin{tabular*}{\textwidth}{@{\extracolsep{\fill}}lccrccc}
\toprule
Research team & Core members & Non-core members & Research field & Team projects & Published papers & Paper citations \\
\midrule
Team 1 & 3 & 8  & Data Mining            & 16 & 413 & 573  \\
Team 2 & 3 & 10 & Machine Learning       & 24 & 672 & 1041 \\
Team 3 & 4 & 8  & Information Retrieval  & 13 & 407 & 639  \\
Team 4 & 5 & 12 & Information Retrieval  & 21 & 625 & 841  \\
Team 5 & 4 & 19 & Data Mining            & 29 & 701 & 1226 \\
\bottomrule
\end{tabular*}
\end{strip}

NMI is 0.81, 0.86, 0.84, and 0.86 for the top 5, 10, 15, and 20 teams, respectively. These results show that AHinRTI effectively identifies scientific research teams.

\clearpage
\section{Conclusions}

By building an academic heterogeneous information network, this paper addresses the inability of homogeneous networks to accurately represent diverse scientific and technological data with complex structures and implicit semantics. AHinE learns low-dimensional, dense, real-valued node representations while preserving rich topology and meta-path semantics. Node-level attention evaluates influence, and AHinRTI uses the learned representations and influence rankings to discover scientific research teams and important members. Experimental results show that the proposed method outperforms the comparison methods.

\begin{acks}
This work was supported by the National Natural Science Foundation of China (62192784, U22B2038, and 62172056).
\end{acks}

\printbibliography[title={References}]

\end{document}